# Robust and Symmetric Magnetic Field Dependency of Superconducting Diode Effect in Asymmetric Dirac Semimetal SQUIDs

H.C. Travaglini[1], J.J. Cuozzo[1], K.R. Sapkota[2], I.A. Leahy[3], A.D. Rice[3], K. Alberi[3], W. Pan[1]

[1] Sandia National Laboratories, Livermore, California 94551, USA

[2] Sandia National Laboratories, Albuquerque, New Mexico 87123, USA

[3] National Renewable Energy Laboratory, Golden, Colorado 80401, USA

Abstract:

The recent demonstration of the superconducting diode effect (SDE) has generated renewed interests in superconducting electronics in which devices such as compact superconducting diodes that can perform signal rectification where low-energy operations are needed. In this article, we present our results of robust and symmetric-in-magnetic-field SDE in asymmetric superconducting quantum interference devices (SQUIDs) realized in high-quality Dirac semimetal $Cd_3As_2$ thin film grown by the molecular beam epitaxy (MBE) technique. Consistent with previous work, a zero magnetic field SDE is observed. Furthermore, the difference in switching current is independent of the strength and polarity of an out-plane magnetic field in the range of -10 mT and 10 mT. We speculate that this robust symmetric-in-field SDE in our Dirac semimetal SQUIDs is due to the formation of helical spin texture, theoretically predicted in Dirac semimetals.

I.  **Introduction**

Applications of information and communications technology have dramatically increased power consumption in recent years. In 2014, data centers worldwide consumed around 194 terawatt hours of electricity, ~ 1% of total demand. It is expected that by 2030 the electricity uses for information and communications technology may exceed 20% of all global electricity [1]. To sustain this explosive development, low-energy, power efficient electronic devices must be developed. Superconducting electronics [2,3] has been a promising platform for low-energy, power efficient microelectronics. Yet, in order for superconducting electronics to reach the level of integration like in silicon microelectronics, devices such as compact superconducting diodes that can perform signal rectification and are compatible with low-energy operations are needed.

The recent demonstration of the superconducting diode effect (SDE) [4], which occurs in certain superconducting materials and their based Josephson junctions (JJs) when both inversion symmetry and time-reversal symmetry (TRS) are broken [5,6], may be able to fulfill this need. In a superconducting diode, the switching (or critical) currents $I_{\pm c}$ are different if the current is swept in different directions (i.e., from zero to a positive (+) value or to negative (-)). For the sake of discussion, here we assume $I_{-c} > I_{+c}$. Consequently, for a current pulse of square wave, if its amplitude ($I_0$) lies in the regime of $I_{-c} > I_0 > I_{+c}$, the output voltage can only assume zero or a finite positive value, i.e., a signal rectification being achieved. These superconducting diodes are useful in offering nearly dissipation free rectification of electrical signals [7,8] and can provide efficient and compact bias-controlled systems needed in superconducting electronics and detectors [9].

It is desired that superconducting diodes can perform robustly over a wide range of magnetic field strengths and directions. This is due to the ubiquitous presence of stray magnetic fields in a superconducting circuit [10,11]. However, most of the superconducting diode effects reported so far are induced by applying a finite magnetic field [12-15], which breaks TRS. As a result, they have different polarities for the different magnetic field directions, i.e., the diode efficiency $\eta = \frac{I_{+c} - I_{-c}}{I_{+c} + I_{-c}}$ displays an anti-symmetric magnetic field dependency. This anti-symmetric field dependency can become problematic in the presence of stray magnetic fields (~ 10 mT in amplitude in a typical superconducting circuit) [10,11]. Indeed, a superconducting diode may work well if a stray field points to one magnetic-field direction and the diode efficiency is positive, but it can stop functioning when the stray field switches its sign, and the diode efficiency becomes negative.

Recently, pioneering work on multiferroic JJs [10] demonstrated that the diode efficiency displays a symmetric-in-field SDE, i.e., independent of magnetic field directions, when helimagnetism is present in multiferroic materials. However, the multiferroic JJs are fabricated using mechanical exfoliation followed by tear-and-stack methods. Thus, they are incompatible with the state-of-the-art semiconductor processing technology. In this article, we present our results of robust and symmetric-in-magnetic-field SDE in asymmetric superconducting quantum interference devices (SQUIDs) on high quality, molecular beam epitaxy (MBE) grown Dirac semimetal $Cd_3As_2$ thin films [16,17]. Consistent with our previous work [18], a zero magnetic field SDE [18-23] is observed in these asymmetric SQUIDs. Furthermore, the difference between the switching currents $I_{\pm c}$ is independent of the strength and polarity of the out-plane magnetic fields in the range of -10 mT and 10 mT. We speculate that this robust, symmetric-in-field SDE in our Dirac semimetal is due to the formation of helical spin texture, theoretically predicted in Dirac semimetals [24-28]. We further point out that a robust and symmetric SDE in SQUIDs using MBE grown $Cd_3As_2$ thin films should allow future superconducting diode applications to take advantage of state-of-the-art semiconductor synthesis and processing.

## II. Experiments

MBE is utilized to grow our high mobility $Cd_3As_2$ thin film of 240 nm thickness. Details about growth can be found in Ref. [16]. For the $Cd_3As_2$ thin film we examine in this work, a semi-insulating GaAs (001) substrate is used. On top of this substrate, a GaAs buffer layer of 500 nm thick is grown at the temperature of 585 °C. It is then followed by a ZnTe (~ 20 nm) and a $Zn_xCd_{1-x}$tep (~ 125 nm) buffer layer, grown at the temperatures of 330 and 290 °C, respectively. Finally, the 240-nm $Cd_3As_2$ layer is grown at 115 °C. A schematic of the growth stack is shown in Fig. 1a. Electronic transport measurements have been carried out in $Cd_3As_2$ thin film of similar growth structures and parameters, and the results are published in Refs. [29,30]. The results there confirm the high quality of the $Cd_3As_2$ thin film.

Electron beam lithography is used to define SQUID structures. Photoresist of 950A6 PMMA is used. After developing the patterns in 1:3 MIBK:IPA, an $O_2$ descum is performed at 100 W for 2 minutes. No other surface treatments, such as argon milling, are performed prior to aluminum (Al) deposition. Electron beam evaporation is used to deposit 5 nm of titanium (to promote metal adhesion to the substrate), 230 nm of Al, and 10 nm of gold (to mitigate oxide formation on the Al). Finally, the photoresist is stripped in a pyrrolidinone-based solution (PG remover). Four

SQUIDs of different sizes are fabricated on one chip. Two SQUID devices are measured. Zero-field SDE is observed in both SQUIDs. Results in this article are presented from one SQUID.

Figure 1b shows an SEM image of this SQUID. The dimension of the SQUID is ~ $1.3 \times 1.0$ μm$^2$. The size of the narrow junction is ~ 150 nm in length and ~ 230 nm in width. For the wide junction, the length and width are ~ 150 and ~ 500 nm, respectively. We note here that $Cd_3As_2$ thin films (devices) when not being measured are stored in a nitrogen purge cabinet prior to (after) device fabrication.

Electronic transport measurements are carried out in a top loading $^3$He system with a base temperature (T) of ~ 300 mK. To minimize the contact resistance, a quasi-four-terminal configuration is used for electronic transport measurements. In this configuration, two gold wires are attached to each superconducting electrode (Fig. 1b) using silver epoxy. A low frequency (~ 11Hz) phase lock-in amplifier (SRS SR830) technique is used to measure the sample resistance $R_{xx}$. The amplitude of AC current excitation ($I_{ac}$) is 300 nA. At this current value, we observe no self-heating effect within our experimental resolution, while the signal to noise ratio is decently large. The current-voltage ($I_{dc}$-$V_{dc}$) measurements are carried out by directly measuring the DC voltage drop ($V_{dc}$) between the two superconducting electrodes. A Keithley 238 source meter is used to provide $I_{dc}$.

### III. Results

Figure 1c shows the quasi-four-terminal sample resistance $R_{xx}$ as a function of temperature (T). The resistance is nearly constant and reaches $R_{xx}$ ~ 6 Ω at high temperatures. It drops to zero at $T_c \sim 1.2$ K signaling a superconducting transition in the device. Based on this transition temperature, a superconducting energy gap $\Delta = 3.5\, k_B T_c = 200$ µeV is obtained. Here $k_B$ is the Boltzmann constant. We point out here that the proximity induced superconducting state has been reported in CVD grown $Cd_3As_2$ nanobelts [31], mechanically exfoliated $Cd_3As_2$ thin flake [32,33], and MBE grown $Cd_3As_2$ in which $Cd_3As_2$ is a 2D topological insulator [34]. Our result here shows that superconductivity can also be induced by the proximity effect in thick MBE grown $Cd_3As_2$ films in which $Cd_3As_2$ is a Dirac semimetal [16].

The inset of Fig. 1c shows $R_{xx}$ as a function of magnetic (B) field measured at 0.3 K. In the low field regime |B| < 50 mT, the device is in the supercurrent regime and $R_{xx} = 0$ Ω. Beyond a critical magnetic field of ~ 53.5 mT, the supercurrent state is destroyed, and the device enters the normal state regime. Here, $B_c$ is determined by taking the B field value at which $R_{xx}$ is 0.6 Ω, 10% of the normal state resistance, a purely empirical criterion also adopted by others [35].

In Fig. 1d, we compare the temperature dependence of $R_{xx}$ obtained from two SQUIDs with similar size; one made of MBE grown $Cd_3As_2$ and the other made of exfoliated $Cd_3As_2$ [18]. To facilitate this comparison, $R_{xx}$ and T are normalized by the normal state resistance and a characteristic temperature $T_0$ (roughly equal to the temperature at which $R_{xx}$ becomes non-zero), respectively. Two features are worthwhile pointing out. First, the normal state resistance of the SQUID examined in this work is much smaller. Second, the superconducting transition is sharper in the current SQUID. Both features demonstrate that a high-quality MBE-grown $Cd_3As_2$ thin film and a high quality SQUID have been achieved. Moreover, it is interesting to notice that there exists an overshot of resistance in the MBE SQUID before the superconducting transition, while in the exfoliation SQUID $R_{xx}$ decreases monotonically. Similar overshot anomaly has also been observed in other materials [35-38]. In a comprehensive study reported in Ref. [35], this resistance anomaly is explained by the different spatial gradients of the quasiparticle and pair electrochemical potentials in the superconductor. In $Cd_3As_2$, the existence of both surface and bulk channels may cause nonequilibrium charge imbalance. Yet, this charge imbalance should occur in both MBE and exfoliation SQUIDs. In other words, an overshot in resistance should also be expected in the exfoliation SQUID, which disagrees with the experimental observation in Ref. [18]. In Refs. [36] and [37], this anomaly is attributed to a decrease in the density of state near $E_F$ when Cooper pairs are formed in the presence of the superconducting proximity effect. Again, this mechanism should apply to both MBE and exfoliation SQUIDs. More studies are needed to understand the overshot anomaly in our MBE SQUID.

Figure 2a shows the I-V characteristics in two current sweep directions in the MBE grown $Cd_3As_2$ SQUID, measured at zero magnetic (B) field. First, we sweep current from 0 to ~ +70 µA, and label this sweep as 0→P. $V_{dc}$ is zero when the current is low. It becomes non-zero at a current of 43.3 µA. We label this switching current as $I_{+c}$. Here, we determine the value of $I_{+c}$ when $V_{dc}$ is equal to 20 µV. In the second sweep, the current varies from 0 to ~ -70 µA. We label it as 0→N. A switch from superconducting state to normal state is again observed at $-I_{-c}$ = - 44.7 µA (at which $|V_{dc}|$ = 20 uV). With the obtained $I_{+c}$ ($I_{-c}$), the value of $eI_{+c}R_n$ ($eI_{-c}R_n$) is calculated to be ~ 260 (270) µeV. They are higher than the Al superconducting gap of $\Delta$ = 200 µeV, demonstrating a high interface transparency [39] between the Al contacts and the $Cd_3As_2$ thin film.

Importantly, $I_{+c} \neq I_{-c}$ demonstrates that the superconducting diode effect has been observed in our asymmetric SQUID at zero magnetic field. This result is consistent with our previous results in a similar asymmetric SQUID but made in a mechanically exfoliated $Cd_3As_2$ thin flake [18]. This shows that the zero-field SDE in $Cd_3As_2$ is an intrinsic behavior and independent of how the material is prepared.

To reveal the SDE more clearly, in Figure 2b, we plot the 0→N trace together with the 0→P trace. The absolute value of both the current and voltage in the 0→N trace is used. It can be seen clearly that in both the supercurrent (where $V_{dc} = 0$) and normal states (where I-V is linear), the two I-V curves overlap perfectly with each other. On the other hand, they differ significantly in the switching regime, revealing the SDE effect.

Temperature dependence is investigated at higher temperatures at T = 0.7, and 0.9 K. It is clearly seen that the diode effect decreases with increasing temperatures. At 0.9K, the diode effect basically disappears (Fig. 3c). We note here that in this device the SDE disappears before the superconducting transition temperature $T_c$ of 1.2K. This seems to suggest that the superconducting surface-bulk coupling, the mechanism that gives rise to broken TRS and thus zero-field SDE in $Cd_3As_2$ [18], may be prone to the order-parameter fluctuations near $T_c$ and is only stable below $T_c$. At $T > T_c$, the device is in the normal state and the I-V curve displays a perfect linear dependence.

In the following, we present our main result of robust, symmetric-in-magnetic-field SDE. All the measurements are carried out at the base temperature of 0.3K. In Figs. 4a and 4b, we show two exemplar I-V curves (including both 0→P and 0→N) at B = 10 and -15 mT, respectively. It is clearly seen that at B = 10 mT, the difference in critical currents is about as same as that at B = 0 mT. The difference deceases when |B| is further increased. At B = -15 mT, $I_{-c} - I_{+c}$ = 0.9 µA. At even higher B fields when the superconductivity is destroyed, $I_{-c} - I_{+c}$ = 0 and the SDE disappears. Fig. 4c shows the switching current $I_{+c}$ and $I_{-c}$ as a function of B. We note here that the measurements at B = 0 mT are repeated before and after the field dependent measurements. The two sets of results are consistent with each other, demonstrating the reproducibility of the SDE. In Fig. 4d, $I_{-c}-I_{+c}$ is plotted against the B field. Within the field range of -10 mT ≤ B ≤ 10 mT, $I_{-c}-I_{+c}$ is nearly constant, ~ 1.5 µA.

### IV. Discussions

The observation of a zero-field SDE is consistent with our previous work on mechanically exfoliated $Cd_3As_2$ thin flake [18], in which we show that the coupling between the surface and bulk superconducting channels in a Dirac semimetal can provide an intrinsic TRS breaking mechanism and, consequently, give rise to a zero-field SDE [18]. We believe that the same mechanism can be applied here for the observed zero-field SDE in the SQUIDs made of MBE grown $Cd_3As_2$ thin films.

In the following, we present in detail the two-band mechanism for zero-magnetic-field SDE within the Ginzburg-Landau theory where the free energy of a JJ is given by

$$F = F_0 - \Gamma_1 \Delta_{L1}^* \Delta_{R1} - \Gamma_2 \Delta_{L2}^* \Delta_{R2} - \frac{\Gamma'}{2}((\Delta_{L1}^* \Delta_{R1})^2 + (\Delta_{L2}^* \Delta_{R2})^2) - \gamma(\Delta_{L1}^* \Delta_{L2} + \Delta_{R1}^* \Delta_{R2}) + \text{c.c.}, \quad (1)$$

where $F_0$ is the free energy of the normal state, $\Delta_{Si}$ are the superconducting order parameters of the two bands $i = 1,2$ at the left ($S = L$) and right ($S = R$) sides of the JJ. $\Gamma_{1(2)}$ describes intra-band Cooper pair tunneling across the JJ, $\Gamma'$ describes pair co-tunneling, and $\gamma$ describes interband pair tunneling. Each order parameter can be written as $\Delta_{Si} = \Delta_i e^{i\varphi_{Si}}$ for $S = L, R$ and $i = 1,2$. Then $F$ can be expressed as

$$F = F_0 - F_1 \cos \varphi_1 - F_2 \cos \varphi_2 - \frac{1}{2}(F'_1 \cos 2\varphi_1 + F'_2 \cos 2\varphi_2) - f_{12}(\cos \varphi_R + \cos \varphi_L), \quad (2)$$

where $\varphi_i = \varphi_{Ri} - \varphi_{Li}$, $\varphi_S = \varphi_{S2} - \varphi_{S1}$, $F_i = \Gamma \Delta_i^2$, $F_i = \Gamma \Delta_i^2$, $F'_i = \Gamma' \Delta_i^4$, and $f_{12} = \gamma \Delta_1 \Delta_2$. Following Ref. [33], we set $\varphi_R = -\varphi_L$ and perform a change of variables: $\varphi_{1/2} = \theta \pm \psi$. Then the supercurrent in the JJ is given by $I_s = \frac{\Phi_0}{2\pi} \frac{\partial F}{\partial \theta}$ where $\Phi_0$ is the superconducting magnetic flux quantum, resulting in a current-phase relationship (CPR)

$$I_s(\theta, \psi) = I_1[(1 + i_2) \cos \psi \sin \theta + (1 - i_2) \sin \psi \cos \theta + (i'_1 + i'_2) \cos 2\psi \sin 2\theta +$$
$$(i'_1 - i'_2) \sin 2\psi \cos 2\theta]. \quad (3)$$

Here $I_1 = F_1 \Phi_0/(2\pi)$, $i_2 = F_2/F_1$, and $i'_{1,2} = F'_{1,2}/F_1$. We see that an equilibrium phase difference $\psi = \pm \pi/2$ breaks TRS in Eq. (3) but does not lead to a zero-field SDE, as observed in $Cd_3As_2$-based JJs [32].

In a SQUID setup, the supercurrent is given by a linear combination of the CPRs in the two arms of the SQUID (labeled $a$ and $b$): $I_s^{SQUID} = I_a(\theta_a, \psi_a) + I_b(\theta_b, \psi_b)$. Fluxoid quantization of the SQUID gives us $\theta_a - \theta_b = \Phi$ where $\Phi = \Phi_{ext}/\Phi_0$ is the normalized flux threading the SQUID ring. Defining $\Theta = (\theta_a + \theta_b)/2$, we have $I_s^{SQUID}(\Theta, \Phi) = I_a\left(\Theta + \frac{\Phi}{2}, \psi_a\right) + I_b\left(\Theta - \frac{\Phi}{2}, \psi_b\right)$. In equilibrium, following our previous work [18], $\psi_a = -\psi_b$ and $0 < |\psi_a| < \pi/2$. In this antiferromagnetic-like SQUID ground state, Eq. (3) implies the zero field SDE is suppressed when the two arms are symmetric [33]. On the other hand, in asymmetric SQUIDs, as examined in this study, a zero field SDE is found.

While our two-band model can provide a mechanism for a SDE at zero magnetic field, the field resilience of our devices requires an analysis going beyond this model. At present, the exact origin for the field resilience SDE in our $Cd_3As_2$ SQUID is not known. Here, we propose a possible mechanism. To this end, we refer to recent work on SDE in multiferroic Josephson junctions and speculate that an antiferromagnetic spin-spiral order (helimagnetism) might be responsible for this resilient symmetric-in-field SDE behavior. Indeed, an antiferromagnetic phase has been predicted in Dirac semimetals like $Cd_3As_2$ in the presence of interactions [24-28,40]. Under this picture, a Dirac semimetal is first viewed as layered quantum spin Hall (QSH) insulators [24]. In each QSH layer, due to edge disorder in realistic samples, a fraction of the bulk electrons can become localized close to the device edge [25] and provide spin-1/2 local moments. These localized spins interact with the electrons of helical edge channels. Consequently, the formation of helical spin textures can occur [24-28]. In another scenario, nuclear spins of the bulk host material can provide a spiral order [26]. In either case, helimagnetism can form in $Cd_3As_2$ and could serve as a possible explanation for a robust SDE.

### V. Conclusion

In this work, we demonstrate a zero magnetic field SDE in asymmetric SQUIDs based on high-quality MBE grown $Cd_3As_2$ thin films. We further show that the coupling between the surface and bulk superconducting channels together with the asymmetry between the two Josephson junction arms in the SQUIDs, a mechanism originally proposed in Ref. [18], are responsible for the zero-field SDE. Moreover, the observed SDE is robust and displays a symmetric-in-field dependence in the field range of $-10 \leq B \leq 10$ mT. We speculate that a helimagnetic order in $Cd_3As_2$, which was theoretically predicted to exist in the presence of interactions [24-28], might be responsible for the observed resilient, symmetric-in-field SDE behavior. Finally, we point out that our results demonstrate that aluminum can form a high-quality interface with as-grown MBE grown $Cd_3As_2$ thin films. This should allow future superconducting electronics applications to take advantage of state-of-the-art semiconductor synthesis and processing.

**Acknowledgements**

We thank Mr. A. Gashi for the help with SEM. The work at Sandia is supported by the LDRD program. W.P. also acknowledges support from DOE, Grant No. DE-SC0022245 for data analysis. Device fabrication was performed at the Center for Integrated Nanotechnologies, a U.S. DOE, Office of BES, user facility. Part of work was carried out at the Molecular Foundry, which is supported by the Office of Science, Office of Basic Energy Sciences, of the U.S. Department of Energy under Contract No. DE-AC02-05CH1123. Sandia National Laboratories is a multi-mission laboratory managed and operated by National Technology & Engineering Solutions of Sandia, LLC (NTESS), a wholly owned subsidiary of Honeywell International Inc., for the U.S. Department of Energy's National Nuclear Security Administration (DOE/NNSA) under contract DE-NA0003525. This written work is authored by an employee of NTESS. The employee, not NTESS, owns the right, title and interest in and to the written work and is responsible for its contents. Any subjective views or opinions that might be expressed in the written work do not necessarily represent the views of the U.S. Government. The publisher acknowledges that the U.S. Government retains a non-exclusive, paid-up, irrevocable, world-wide license to publish or reproduce the published form of this written work or allow others to do so, for U.S. Government purposes. The DOE will provide public access to results of federally sponsored research in accordance with the DOE Public Access Plan. This work was authored in part by the National Renewable Energy Laboratory, operated by Alliance for Sustainable Energy, LLC, for the U.S. Department of Energy (DOE) under Contract No. DE-AC36-08GO28308. Funding for $Cd_3As_2$ film growth was provided by the U.S. Department of Energy, Office of Science, Basic Energy Sciences, Division of Materials Sciences and Engineering, Physical Behavior of Materials Program under the Disorder in Topological Semimetals project.

Figures

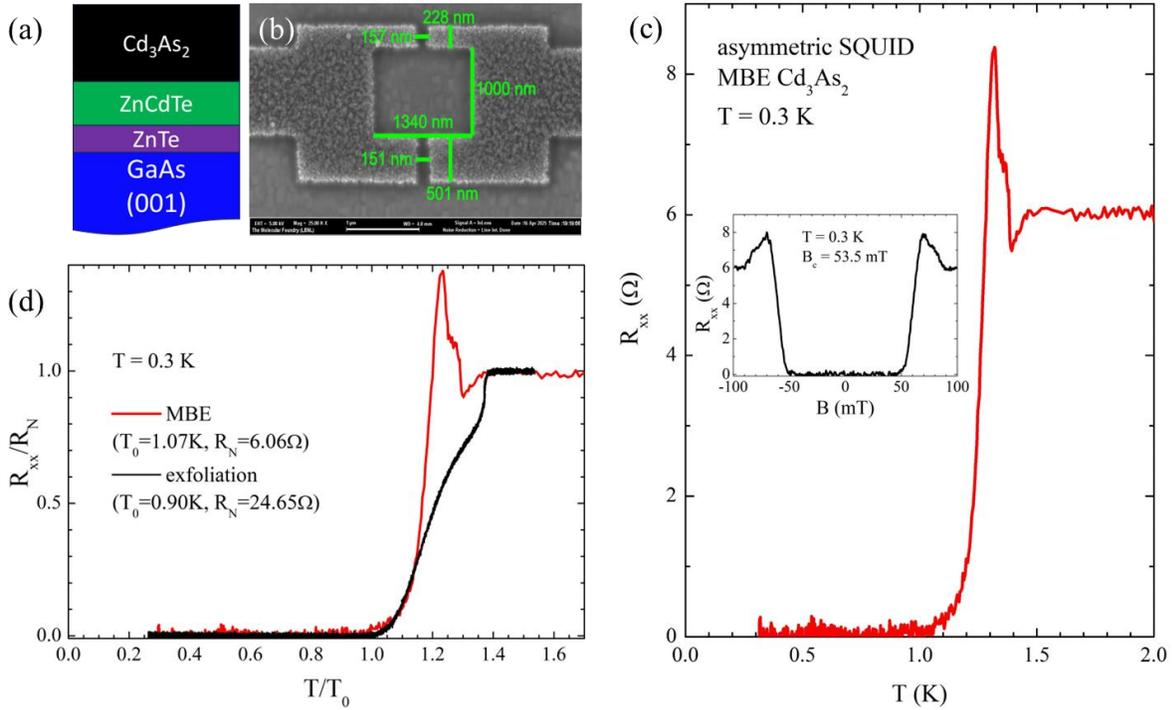

Figure 1: Superconducting transition in an asymmetric SQUID in an MBE grown $Cd_3As_2$ thin film. (a) shows a schematic of the growth stack of the MBE grown $Cd_3As_2$ film. (b) shows an SEM image of the asymmetric SQUIDs studied. The gray background represents the $Cd_3As_2$ thin film. (c) shows the quasi-four-terminal sample resistance $R_{xx}$ as a function of temperature (T). A superconducting transition is observed at $T_c \sim 1.2K$. The inset displays the magnetic (B) field dependence of $R_{xx}$, measured at T = 0.3K. From this magnetoresistance, a critical magnetic field of ~ 53.5 mT is obtained. (d) shows the T dependence of $R_{xx}$ in both the MBE grown (this work) and mechanically exfoliated (Ref. [18]) $Cd_3As_2$ based SQUIDs. It is clearly seen that the superconducting transition is sharper in the MBE SQUID. Moreover, a resistance overshot is observed in the MBE SQUID.

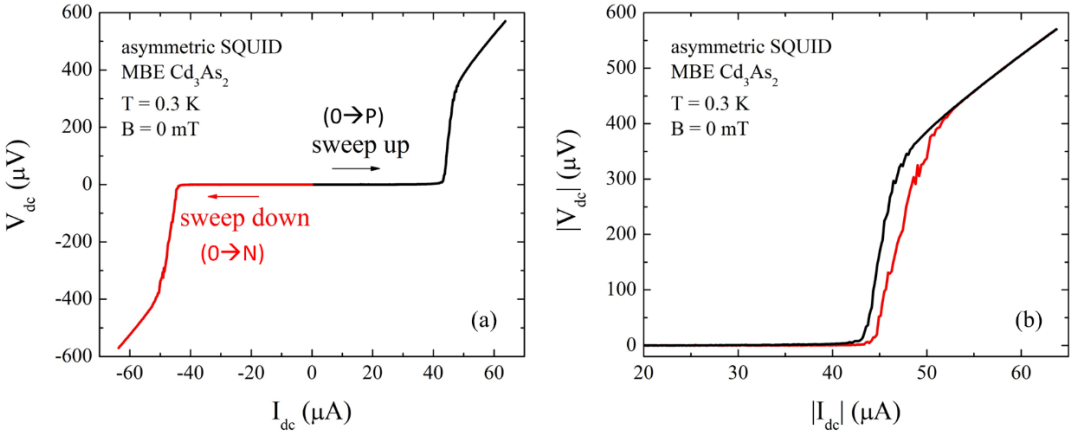

Figure 2: Zero magnetic field superconducting diode effect. (a) Current-voltage (I-V) characteristics measured at $B = 0$ mT. 0→P and 0→N refer to the curves in which the current sweeps from 0 to ~ + 70 µA and 0 to ~ − 70 µA, respectively. In (b), the absolute value of the 0→N curve is plotted along with the 0→P curve. The two curves overlap perfectly in both the superconducting and normal state regimes, but differ clearly in the switching regime, demonstrating the zero-field superconducting diode effect.

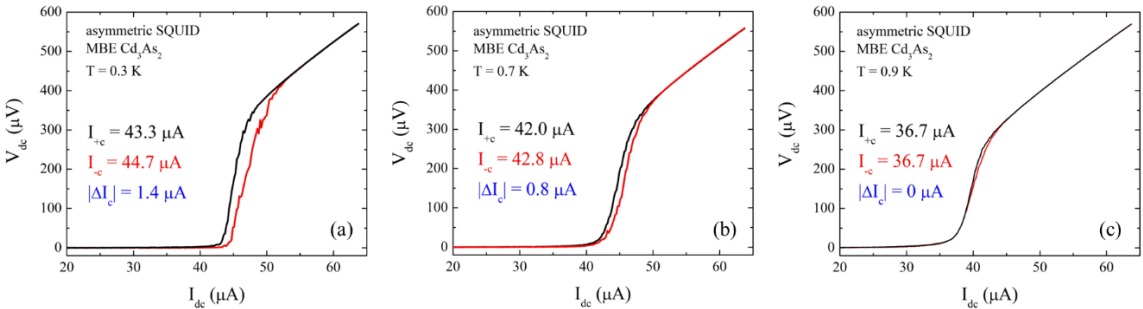

Figure 3: Temperature depedence of the zero-field superconducting diode effect. I-V characteristics at zero magnetic field measured at three temperatures, 0.3K (a), 0.7 K (b), and 0.9 K (c). It is clearly seen that the diode effect decreases with increasing temperatures.

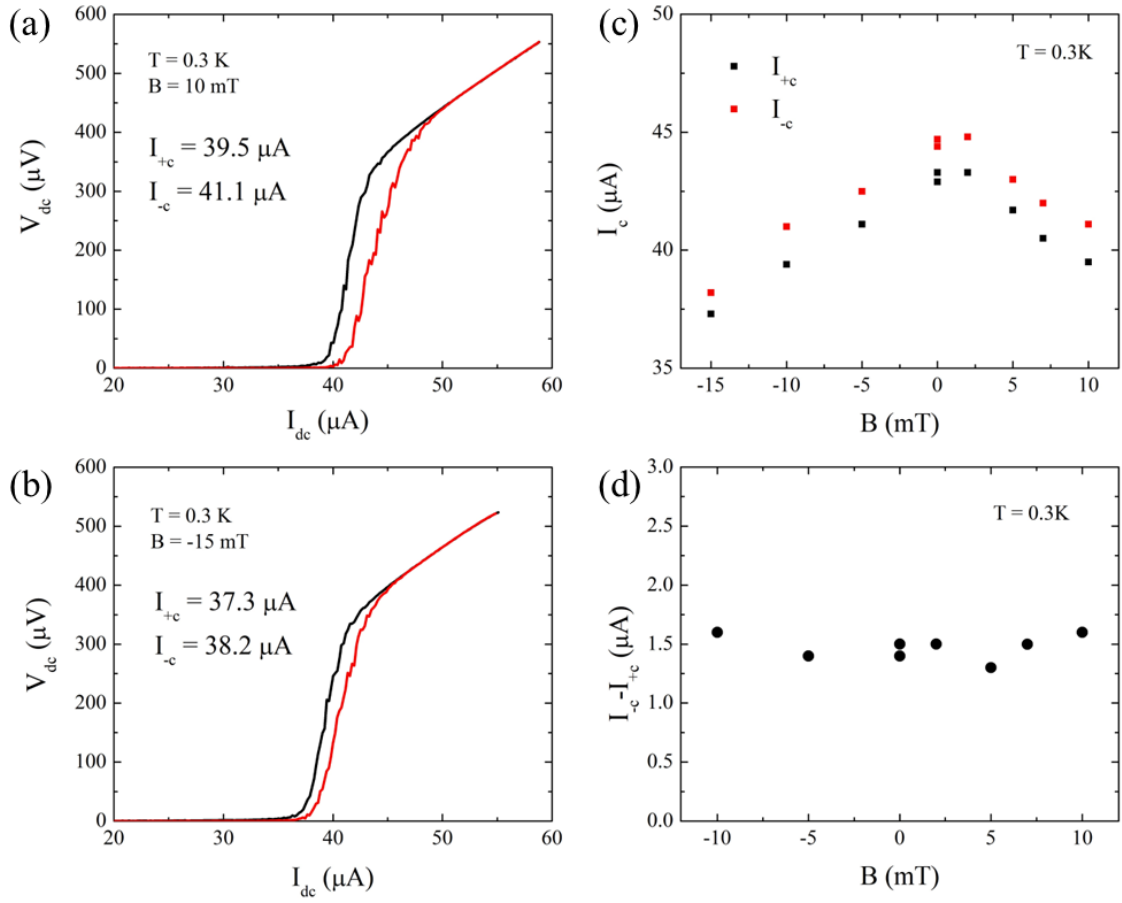

Figure 4: Magnetic field depedence of the superconducting diode effect. (a) and (b) show the I-V characteristics at two magnetic fields of different polarity: 10 mT (a) and -15 mT (b). (c) shows the magnetic (B) field dependence for both $I_{+c}$ and $I_{-c}$. In (d), the difference of critical current, $I_{-c} - I_{+c}$, is plotted as a function of B field. In the field range of $-10 \leq B \leq 10$ mT, $I_{-c} - I_{+c}$ is nearly constant, ~ 1.5 µA.